\documentclass[pre,aps,preprint,showpacs,floatfix,showkeys,byrevtex]{revtex4}
\usepackage[dvips]{graphicx}
\usepackage{amsmath}
\usepackage{amsfonts}
\usepackage{amssymb}

\begin{document}

\title{Implication of Barrier Fluctuations on the Rate of Weakly Adiabatic Electron Transfer}

\author{Bart{\l}omiej \surname{Dybiec}}
\email{bartek@th.if.uj.edu.pl}

\author{Ewa \surname{Gudowska-Nowak}} 
\email{gudowska@th.if.uj.edu.pl}

\author{Pawe{\l} F. \surname{G\'ora}} 
\email{gora@if.uj.edu.pl}

\affiliation{Marian~Smoluchowski Institute of Physics,\\
 Jagellonian University, Reymonta~4, 30--059~Krak\'ow, Poland}

\date{\today}

%
%
\begin{abstract}
The problem of escape of a Brownian particle in a cusp-shaped metastable potential is of
special importance in nonadiabatic and weakly-adiabatic rate theory for electron transfer (ET)
reactions. Especially, for the weakly-adiabatic reactions, the reaction follows an adiabaticity
criterion in the presence of a sharp barrier. In contrast to the non-adiabatic case, the ET
kinetics can be, however considerably influenced by the medium dynamics.\\
In this paper, the problem of the escape time over a dichotomously fluctuating cusp barrier
is discussed with its relevance to the high temperature ET reactions in condensed media.
\end{abstract}

\vspace*{5pt}


\maketitle

\section{Introduction}	
\begin{figure}[!h]
\includegraphics[angle=0, width=8.5cm, height=8.5cm]{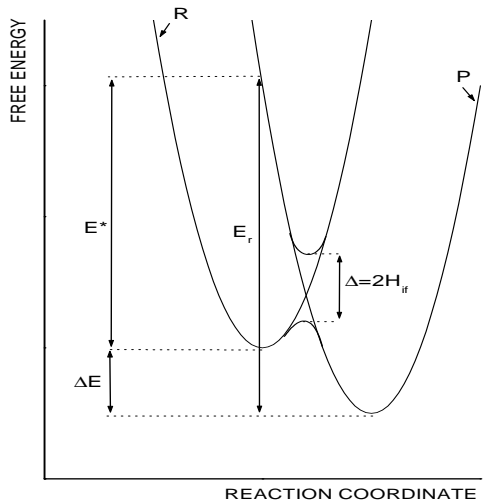}
\caption{\label{et_co}
Schematic energy profiles of reactant ({\bf R}) and product ({\bf P})
states of the electron transfer reaction coordinate. The reorganization energy
$E_r$ is the sum of the reaction energy $\Delta E$  and the optical excitation
energy $E^*$. $\Delta$ stands for energy separation between the energy surfaces
due to electronic coupling of {\bf R} and {\bf P} states. Thermal ET occurs at
nuclear configurations characteristic to the intersection of the parabolas.}
\end{figure}

Mechanism of the electron transfer (ET) in condensed and biological media goes beyond universal
nonadiabatic approach of the Marcus theory.\cite{Marcus1,Marcus2,Ulstrup,Kuznetsov,Chandler,Makarov} 
In particular, relaxation properties of medium 
may slow down the
overall ET kinetics and lead to an adiabatic dynamics.\cite{Hynes} An excess electron appearing in the
medium introduces local fluctuations of polarization, that in turn contribute to the
change of Gibbs energy. Equilibration of those fluctuations leads to a new state with a
localized position of a charge.
In chemical reactions, the electron may change its location passing from a donoring to an
accepting molecule, giving rise to the same scenario of Gibbs energy changes that allows to
discriminate between the (equilibrium) states ``before'' and ``after'' the transfer 
(see Fig.~\ref{et_co}). 
The free energy surfaces for ``reactants'' and ``products'' are usually multidimensional functions
which intersect at the transition point. The deviation from it, or the Gibbs energy change, 
can be calculated from the reversible work done along the path that forms that state,
so that by use of a simple thermodynamic argument, one is able to associate a change in the
Gibbs energy with the change of multicomponent ``reaction coordinate'' that describes a
response of the system to the instantaneous transfer of a charge from one site to another.

%
%

ET reactions involve both classical and quantum degrees of freedom. Quantum effects are mostly
related to electronic degrees of freedom. Because of the mass difference between electrons and
nuclei, it is frequently assumed that the electrons follow the nuclear motion adiabatically
(Born-Oppenheimer approximation). The interaction between two different electronic states
results in a splitting energy $\Delta$. The reaction from reactants to products is then mediated by
an interplay of two parameters: time of charge fluctuations between the two neighbouring
electronic states and a typical time  within which the nuclear reaction coordinate crosses the
barrier region. When the electronic ``uncertainty'' time of charge fluctuations is shorter than 
the nuclear dynamics, the transition is adiabatic with the overall dynamics evolving on an
adiabatic ground-state surface. For small splitting between electronic states $\Delta\approx 0.1-1$, this
ground-state adiabatic surface is often characterized by a cusp-shaped 
potential.\cite{Marcus1,Ulstrup,Kuznetsov,Hynes} In fact, it is often
argued\cite{Hynes} that majority of natural ET reactions are in what we term
weakly adiabatic regime, when the reaction is still adiabatic but the barrier is
quite sharp and characterized by the barrier frequency $\omega_a$ roughly by an
order of magnitude higher than the medium relaxation frequency $\omega_0$.

The dynamics of the reaction coordinate in either of the potential wells can be estimated by
use of a generalized Langevin equation with a friction term mimicking dielectric response of
the medium. We are thus left with a standard model of a Brownian particle in a (generally
time-dependent) medium.\cite{Hynes,Hynes1}
In the case of a cusp-shaped potential, a particle approaching the top of a barrier with
positive velocity will almost surely be pulled towards the other minimum.
The kinetic rate is then determined\cite{Hynes,Hanggi} by the reciprocal mean first passage
time (MFPT) to cross the barrier
wall with positive velocity $\dot{x}>0$ 
 and in a leading order in the barrier height $\Delta E/k_BT$  yields the standard
transition state theory (TST) result
\begin{equation}
k^{cusp}=\frac{\omega_0}{2\pi}\exp(-\Delta E/k_BT).
\label{cusp}
\end{equation}

As discussed by Talkner and Braun,\cite{Talk} the result holds also for
non-Markovian processes with memory friction satisfying the
fluctuation-dissipation theorem. The TST formula follows from the Kramers rate
for the spatial-diffusion-controlled rate of escape at moderate and strong
frictions $\eta$
\begin{equation}
k_{R \rightarrow
P}=\frac{(\eta^2/4+\omega_a^2)^{1/2}-\eta/2}{\omega_a}\frac{\omega_0}{2\pi}
\exp(-\Delta E/k_BT).
\end{equation}
Here $\omega_a$ stays for the positive-valued  angular frequency of the unstable state at the
barrier, and $\omega_0$ is an angular frequency of the metastable state at $x=R$. For strong
friction, $\eta>>\omega_a$ the above formula leads to a common Kramers result
\begin{equation}
k_{R \rightarrow P}=\frac{\omega_0 \omega_a}{2\pi\eta}\exp(-\Delta E/k_BT),
\end{equation}
and reproduces the TST result eq.~(\ref{cusp}) after letting the barrier frequency tend to
infinity
$\omega_a\rightarrow\infty$ with $\eta$ held fixed. Moreover, as pointed out by Northrup and
Hynes,\cite{Hynes1} in a weakly adiabatic case, the barrier ``point'' is a negligible fraction of the high
energy barrier region, so that the rate can be influenced by medium relaxation in the wells. The
full rate constant for a symmetric reaction is then postulated in the form
\begin{equation}
k_{WA}=\left(1+2k^a_{WA}/k_D\right)^{-1}k_{WA}^a
\end{equation}
where $k_{WA}^a\approx k^{cusp}$ and $k_D$ is the well (solvent or medium) relaxation rate constant which for a harmonic
potential and a high barrier ($\Delta E\ge 5k_BT$) within 10\% of accuracy simplifies to
\begin{equation}
 k_D=\frac{m_0 \omega_0^2 D}{k_BT}\left (\frac{\Delta E}{\pi k_BT}\right)^{1/2}
 e^{-\Delta E/k_BT}.
\end{equation}

In the above equation, the diffusion constant $D$ is related\cite{Ulstrup} {\it via} linear response theory to the
longitudinal dielectric relaxation time $\tau_L$ and for symmetric ET reaction
($m_0^2\omega_0^2=2E_r$) reads
\begin{equation}
D=\frac{k_BT}{m_0\omega_0^2\tau_L}=\frac{k_BT}{2E_r \tau_L}.
\end{equation}

Existence of a well defined rate constant for a chemical reaction requires that the relaxation time
of all the degrees of freedom involved in the transformation, other than the reaction coordinate,
is fast relative to motion along the reaction coordinate. If the separation of time scales were not
present, the rate coefficient would have a significant frequency dependence reflecting various modes
of relaxation. Such a situation can be
expected in complex media, like non-Debeye solvents or proteins, where there are many different
types of degrees of freedom  with different scales of relaxation.  Although in these cases the
 rate ``constant'' can no longer be defined, the overall electron transfer can be described in terms
 of the mean escape time that takes into account noisy character of a potential surface along with
 thermal fluctuations. 
 
 The  time effect of the surroundings (``environmental noises''), expressed by different time constants
 for polarization and the dielectric reorganization 
  has not been so far explored in detail, except in photochemical reaction centers\cite{Chandler}
 where molecular dynamics studies have shown that the slow components of the 
 energy gap fluctuations are, most likely, responsible for the observed nonexponential kinetics
 of the primary ET process. The latter will be assumed here to
 influence the activation free energy of the reaction $\Delta E=E_r/4$ and will be envisioned as
 a barrier alternating processes. In
 consequence,
 even small variations $\delta E$  in $\Delta E$ can
greatly modulate 
the escape kinetics  (passage from reactants' to products' state) in the system.
If the barrier fluctuates extremely slowly, the mean first passage time to
the top of the barrier is dominated  by those realizations for which the
barrier starts in a higher position and thus becomes very long. The barrier
is then essentially quasistatic throughout the process. At the other extreme,
in the case of rapidly fluctuating barrier, the mean first passage time is
determined by the ``average barrier''.
 For some particular correlation time of the barrier
fluctuations, it can happen however, that the mean kinetic rate of the process 
 exhibits an extremum\cite{doe,iwa,bork} 
that is a signature of resonant tuning of the system in response
to the external noise.

In this contribution we present analytical and  numerical results for the escape kinetics over a
fluctuating cusp for the high temperature model ET system with a harmonic potential 
subject to dichotomous fluctuations. We perform our investigations of the average escape time
as a function of the correlation rate of the dichotomous noise in the barrier height at fixed
temperatures.
In particular, we  examine variability of the mean first passage time for different types of
barrier switching\cite{doe,zurcher,boguna} when the barrier changes either between the
``on-off'' position, flips between a barrier or a well, or it varies between two different
heights. By use of a Monte Carlo procedure we determine probability density function (pdf)
of escape time in the system and investigate the degree of nonexponential behavior in the
decay of a primary state located in the reactants' well.

%
%
\section{Generic Model System}		
At high temperatures it is permissible to treat the low frequency
medium modes classically. The medium coordinates are continuous and it is useful
to draw a one-dimensional schematic representation of the system (Fig.~\ref{et_co}) with
the reaction proceeding almost exclusively at the intersection energy. 
As a model of the reaction coordinate kinetics, we have considered an overdamped Brownian particle
 moving in a potential field
between absorbing and reflecting boundaries in the presence of noise
that modulates the barrier height. 
The evolution of the reaction coordinate $x(t)$ is described in terms of the 
Langevin equation
\begin{equation}
\frac{dx}{dt}=-V'(x)+\sqrt{2T}\xi(t)+g(x)\eta(t)= -V'_{\pm}(x)+\sqrt{2T}\xi(t).
\label{lang}
\end{equation}
Here $\xi(t)$ is a Gaussian process with zero mean and correlation
 $<\xi(t)\xi(s)>=\delta(t-s)$ ({\it i.e.} the Gaussian white noise arising
 from the heat bath of temperature $T$),
$\eta(t)$ stands for  a dichotomous (not necessarily symmetric)
 noise taking on one of
two possible values $a_\pm$ and prime means differentiation over $x$. Correlation time 
of the dichotomous process has
been set up to ${1\over 2\gamma}$ with $\gamma$ expressing the flipping frequency
of the barrier fluctuations. 
Both noises are assumed to be statistically independent, {\it i.e.} $<\xi(t)\eta(s)>=0$.
Equivalent to eq.~(\ref{lang}) is a set of the Fokker-Planck equations 
describing evolution of the probability density of finding the particle
at time $t$  at the position $x$ subject to the force 
$-V'_{\pm}(x)=-V'(x)+a_{\pm}g(x)$

\begin{eqnarray}
\partial_t {P}(x,a_\pm,t)& =&  \partial_x  \left[V'_{\pm}(x)+T\partial_x  \right]P(x,a_\pm,t) \nonumber \\
  & -& \gamma P(x,a_\pm,t)+\gamma P(x,a_\mp,t).
\label{schmidr}
\end{eqnarray}

In the above equations time has dimension of $[\mathrm{length}]^2/\mathrm{energy}$ due to a friction
constant that has been ``absorbed'' in the time variable.
We are assuming absorbing boundary condition at $x=L$ and a  reflecting
boundary at $x=0$

\begin{equation}
P(L,a_\pm,t)=0,
\label{bon0}
\end{equation}
\begin{equation}
\left[V'_{\pm}(x) +T\partial_x\right]P(x,a_\pm,t)|_{x=0}=0.
\label{bon}
\end{equation}

\noindent The initial condition

\begin{equation}
P(x,a_+,0)=P(x,a_-,0)=\frac{1}{2}\delta(x)
\end{equation}

\noindent
expresses equal choice to start with any of the two configurations of the
barrier.
The quantity of interest is the mean first passage time 

\begin{eqnarray}
\mathrm{MFPT} & =& \int\limits_0^\infty dt\int\limits_0^L\left[P(x,a_+,t)+P(x,a_-,t)\right]dx \nonumber \\
& = & \tau_+(0)+\tau_-(0)
\end{eqnarray}

\noindent
with $\tau_+$ and $\tau_-$ being MFPT for $(+)$ and $(-)$ 
configurations, respectively. MFPTs $\tau_+$ and $\tau_-$
 fulfill the set of backward Kolmogorov equations\cite{bork} 

\begin{equation}
-{1\over 2}=-\gamma\tau_\pm (x)+\gamma\tau_\mp(x)-{dV_\pm (x)\over dx}{d\tau_\pm (x)\over dx}+T{d^2\tau_\pm (x)\over dx^2} 
\label{mr_uklad}
\end{equation}

\noindent
with the boundary conditions ({\it cf.} eq.~(\ref{bon0}) and~(\ref{bon}))

\begin{equation}
\tau'_{\pm}(x)|_{x=0}=0,
\qquad
\tau_{\pm}(x)|_{x=L}=0.
\end{equation}

Although the solution of (\ref{mr_uklad}) is usually unique,\cite{molenaar}
 a closed, ``ready to use'' analytical formula for the MFPT can be obtained only for the simplest cases of
 the potentials (piecewise linear). More complex cases, like even piecewise 
 parabolic potential $V_\pm$ result in an intricate form of the solution to
 eq.~(\ref{mr_uklad}).
 Other situations require either use of approximation schemes,\cite{rei}
 perturbative approach\cite{iwa} or direct numerical evaluation methods.\cite{rec,gam}
In order to  examine MFPT for various potentials a modified program\cite{musn}
applying general shooting methods has been used.  Part of the mathematical
software has been obtained  from the {\it Netlib} library. 

%
%
\section{Solution and Results}
Equivalent to equation~(\ref{mr_uklad}) is a set of equations
\begin{equation}
\left[\begin{array}{c}
{du(x)\over dx}\\
{dv(x)\over dx}\\
{dp(x)\over dx}\\
{dq(x)\over dx}\\
\end{array}\right]=
\left[\begin{array}{cccc}
0 & 0 & 1 & 0 \\
0 & 0 & 0 & 1 \\
0 & 0 & {{d\over dx}\left[V_+(x)+V_-(x)\right]\over 2T} & {{d\over dx}\left[V_+(x)-V_-(x)\right]\over 2T} \\
0 & {2\gamma\over T} & {{d\over dx}\left[V_+(x)-V_-(x)\right]\over 2T} &  {{d\over dx}\left[V_+(x)+V_-(x)\right]\over 2T}\\
\end{array}\right]
\left[\begin{array}{c}
u(x)\\
v(x)\\
p(x)\\
q(x)\\
\end{array}\right]+
\left[\begin{array}{c}
0\\
0\\
-{1\over T}\\
0\\
\end{array}\right],
\label{muklad}
\end{equation}

\noindent where new variables have been introduced 

\begin{equation}
\left\{\begin{array}{c}
u(x)=\tau_+(x)+\tau_-(x) \\
v(x)=\tau_+(x)-\tau_-(x)
\end{array}\right.,
\qquad
\left\{\begin{array}{l}
{du(x)\over dx}=p(x) \\
{dv(x)\over dx}=q(x)  
\end{array}\right..
\end{equation}

\noindent
Since $u$ does not enter the right-hand side of any of the above equations, the
system can be further converted to 
\begin{equation}
\left[\begin{array}{c}
{dv(x)\over dx}\\
{dp(x)\over dx}\\
{dq(x)\over dx}\\
\end{array}\right]=
\left[\begin{array}{ccc}
0 & 0 & 1 \\
0 & {{d\over dx}\left[V_+(x)+V_-(x)\right]\over 2T} & {{d\over dx}\left[V_+(x)-V_-(x)\right]\over 2T} \\
{2\gamma\over T} & {{d\over dx}\left[V_+(x)-V_-(x)\right]\over 2T} &  {{d\over dx}\left[V_+(x)+V_-(x)\right]\over 2T}\\
\end{array}\right]
\left[\begin{array}{c}
v(x)\\
p(x)\\
q(x)\\
\end{array}\right]+
\left[\begin{array}{r}
0\\
-1/T\\
0\\
\end{array}\right]
\label{mr_6}
\end{equation}
{\it i.e.} it has a form of
\begin{equation}
{d\vec{f}(x)\over dx}=\hat{A}(x)\vec{f}(x)+\vec{\beta}(x).
\label{vecf}
\end{equation}

\noindent
A unique solution to (\ref{vecf}) 
exists\cite{molenaar} and reads
\begin{eqnarray}
\vec{f}(x)&=&\exp\left\{\int\limits_0^x\hat{A}(x')dx'\right\}\vec{f}(0) \nonumber \\
&+& \exp\left\{\int\limits_0^x\hat{A}(x')dx'\right\}\int\limits_0^x \exp\left\{-\int\limits_0^{x'}\hat{A}(x'')dx''\right\}
\vec{\beta}(x')dx' \nonumber \\
&=& {\bf{A}}(x)\vec{f}(0)+{\bf{A}}(x)\vec{B}(x)
= {\bf{A}}(x)\vec{f}(0)+\vec{C}(x)
\label{rozw}
\end{eqnarray}

\noindent
with boundary conditions leading to 
\begin{equation}
\vec{f}(0)=
\left[
\begin{array}{c}
 \tau_+-\tau_- \\
 0 \\
0 \\
\end{array}
\right],
\qquad
\vec{f}(L)=
\left[
\begin{array}{c}
0 \\
p(L) \\
q(L) \\
\end{array}
\right].
\label{abs}
\end{equation}

\noindent
MFPT is the quantity of interest
\begin{equation}
\tau=\tau(0)=u(0),
\end{equation}

\noindent which can be obtained from

\begin{equation}
p(x)={du(x)\over dx},
\end{equation}
and
\begin{equation}
\int\limits_0^Lp(x)dx=u(L)-u(0)=0-u(0)=-u(0),
\end{equation}

\noindent with
\begin{equation}
u(0)=\tau=-\int\limits_0^Lp(x)dx.
\label{mr_7}
\end{equation}

For the parabolic potential $V_+(x)=-V_-(x)={Hx^2\over L^2}\equiv 2E_r x^2$ 
the above procedure leads to

\begin{equation}
\hat{A}(x)=\left[\begin{array}{ccc}
0 & 0 & 1 \\
0 & 0 & {2Hx\over L^2T}\\
{2\gamma\over T} & {2Hx\over L^2T}& 0\\\end{array}\right], 
 \qquad 
\int\limits_0^x\hat{A}(x)dx=x\left[\begin{array}{ccc}
0 & 0 & 1 \\
0 & 0 & {Hx\over L^2T}\\
{2\gamma\over T} & {Hx\over L^2T}& 0\\\end{array}\right],
\end{equation}

\noindent and
\begin{equation}
\tau=\int\limits_0^L\int\limits_0^L\Phi(x,y)dxdy,
\end{equation}

\noindent where

\begin{eqnarray}
\Phi(x,y)& =&  \left[-{L^2H\over 2}{x[\varphi(x)-2]\over\rho(x)} -{LH[\varphi(L)-2]\over4[H^2+\gamma TL^2\varphi(L)]}{4\gamma
L^4T+H^2x^2\varphi(x)\over \rho(x)}  \right. \nonumber \\
& + &\left.  {\sqrt{\rho(L)}\xi(L)H \over 4[H^2+\gamma TL^2\varphi(L)]}{x\xi(x)\over\sqrt{\rho(x)}}\right]  
\times   H\gamma L^2{y[\varphi(y)-2]\over\rho(y)} \nonumber \nonumber \\
& + &  \left[{H^2L^4\gamma y[\varphi(y)-2]\over 2\rho(y)}{x[\varphi(x)-2]\over \rho(x)} 
   -   {H^2y\xi(y)\over 4T\sqrt{\rho(y)}}{x\xi(x)\over\sqrt{\rho(x)}}  \right. \nonumber \\
& + & \left.{4\gamma L^4T+H^2y^2\varphi(y)\over 4T\rho(y)}  {4\gamma L^4T+H^2x^2\varphi(x)\over \rho(x)} \right] 
 \times\theta(y-x),
\end{eqnarray}
\begin{equation}
\rho(x)=H^2x^2+2\gamma L^4T,
\end{equation}
\begin{equation}
\varphi(x)=\exp\left[{{\sqrt{\rho(x)}x\over L^2T}}\right]+\exp\left[{-{\sqrt{\rho(x)}x\over L^2T}}\right]=2\cosh\left[{{\sqrt{\rho(x)}x\over
L^2T}}\right],
\end{equation}
\begin{equation}
\xi(x)=\exp\left[{{\sqrt{\rho(x)}x\over L^2T}}\right]-\exp\left[{-{\sqrt{\rho(x)}x\over L^2T}}\right]=2\sinh\left[{{\sqrt{\rho(x)}x\over
L^2T}}\right].
\end{equation}

%
%

\begin{figure}[!h]
\includegraphics[angle=0, width=8.5cm, height=8.5cm]{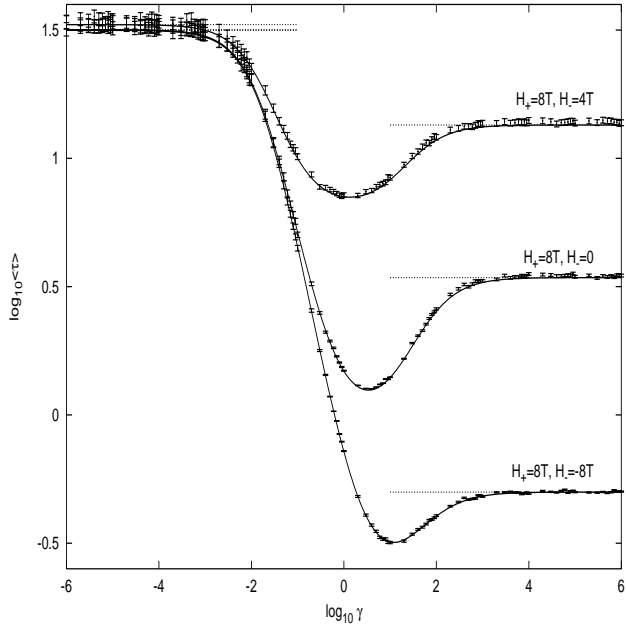}
\caption{\label{flat}
MFPT as a function of the correlation rate of the dichotomous noise
 for parabolic potential barriers switching between different
heights $H_{\pm}$. Full lines: analytical results; symbols stay for the results
from MC simulations with $\Delta t=10^{-5}$ and ensemble of $N=10^4$
trajectories. For simplicity, the parametrization $L=T=1$ has been used.}
\end{figure}

Fig.~\ref{flat} displays calculated MFPT as a function of switching frequency $\gamma$.
Analytical solutions are presented along with results from Monte Carlo simulations of
eq.~(\ref{lang}). As a choice for various configurations of the potential, we have probed
$H_{\pm}=\pm 8T; H_+=8T, H_-=0$ and $H_+=8T, H_-=4T$ that set up reorganization
 energies ($2E_r=H$) and
heights of barrier in the problem of interest.

The distinctive characteristics of resonant activation is observed with the
average escape time initially decreasing, reaching a minimum  value, and then
increasing as a function of switching frequency $\gamma$. At slow dynamics of
the barrier height, {\it i.e.} for values of the rate $\gamma$ less than
$\tau_+^{-1}$, the average escape time approaches the value
$\tau=(\tau_--\tau_+)/2$ predicted by theory\cite{doe,boguna,rei,bier} and observed in experimental
investigations of resonant activation.\cite{Mantegna} For fast dynamics of the
barrier height the average escape time reaches the value associated with an effective
potential characterized by an average barrier. In comparison to the ``on-off''
switching of the barrier, the region of resonant activation
flattens for dichotomic flipping between the barrier and a well, and in the case of
the Bier-Astumian model ($H_+=8T,H_-=4T$) when the barrier changes its height
between two different values. The resonant frequency shifts from the lowest
value for the Bier-Astumian model to higher values for the ``on-off'' and the
``barrier-well'' scenarios, respectively. This observation is in agreement with
former studies\cite{boguna} aimed to discriminate between characteristic
features of resonant activation for models with ``up-down'' configurations
of the barrier and models with the ``up'' configuration but fluctuating between
different heights. The ``up-down'' switching of the barrier heights produces 
shorter MFPT and in consequence, higher value of crossing rates for resonant
frequencies than two other models of barrier switching.

For each of the above situations we have evaluated probability density function
for first escape
times. Pdfs have been obtained as a result of MC simulations  on $N=10^4$
trajectories by use of histograms and kernel
density estimation methods.
In the resonant
activation regime, pdf of escape times has an exponential slope, that suggests
that the reactants' population follows preferentially the kinetics through the
state with the lowest barrier. Similarly, the exponential decay times of
reactants' are observed in the high frequency limit ($\gamma\approx 10^9$), when the system experiences
an effective potential with an average barrier.\cite{doe,zurcher,boguna,rei,gam,Mantegna,reihan,pechukas}

Apparent nonexponential decay of the initial population  
is observed at low frequencies 
($\gamma\approx 10^{-6}$), when the flipping rate becomes less than
$\tau_-^{-1}$. 
%
%
\begin{figure}[!h]
\includegraphics[angle=0, width=8.5cm, height=8.5cm]{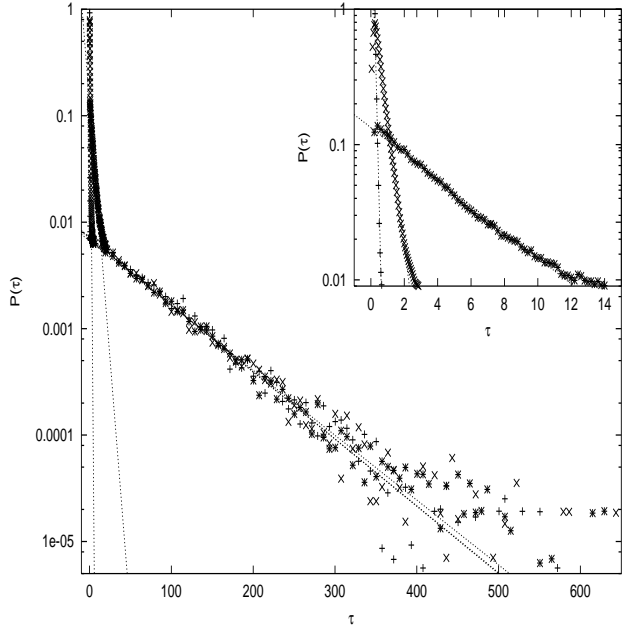}
\caption{\label{scal}Semilog plot of the pdf for first escape times in the system.
The differences between lines fitted to the slopes reflect various $\tau_-$ values 
for a given type of barrier switching: 
($+$) for $H_+=8T,\;H_-=-8T$; ($\times$) for $H_+=8T,\;H_-=0$ and ($\ast$) for $H_+=8T,\;H_-=4T$
({\it cf.} Table~1).}
\end{figure}
As it is clearly demonstrated in Fig.~\ref{scal} and summarized in
Table~1, passage times at low
frequencies are roughly characterized by two distinct time scales that
correspond to $\tau_-$ (different for various switching barrier scenarios - {\it
cf.} first three rows of Table~1) and $\tau_+$ (bottom three rows of Table~1), 
respectively. The inset of Fig~\ref{scal} shows pdf zoomed for part of the main plot, where the
differences between various, model dependent $\tau_-$ values can be well distinguished.

%
%
\begin{table}[h!]
\begin{center}
\begin{footnotesize}
\begin{tabular}{c c c c c } \\
\hline
&   $H_+$ & $H_-$   & fitted value &  static barrier value \\ \hline
& 8T & -8T & $0.09\pm0.01$ & 0.12  \\ 
$\tau_- $& 8T & 0 & $0.46\pm0.01$ & 0.50 \\ 
& 8T & 4T &  $4.51\pm0.05$ & 3.43 \\ \hline

& 8T & -8T & $68.73\pm0.89$ & 63.01 \\ 
$\tau_+$ & 8T & 0 & $71.33\pm3.04$ & 63.01  \\ 
& 8T & 4T & $79.11\pm1.98$ & 63.01 \\ \hline
\end{tabular}
\label{tablel}
\caption{Relaxation time for parabolic potential barriers switching between
different heights $H_\pm$ at low frequencies.}
\end{footnotesize}
\end{center}
\end{table}  
%
%
%

%
%
\section{Summary}
\noindent
In the foregoing sections we have considered the thermally activated
process that can describe classical ET kinetics. The regime of dynamical disorder where
fluctuations of the environment can interplay with the time scale of the reaction itself
is
not so well understood. The examples where such a physical situation can happen are
common to nonequilibrium chemistry and, in particular to ET reactions\cite{Chandler} in
photosynthesis. As a toy model for describing the nonexponential ET kinetics\cite{Chandler} we have chosen
a generic system displaying the resonant activation phenomenon.
The reaction coordinate  has been coupled to an external noise source that can describe
polarization and depolarization processes responsible for the height of barrier between
the reactants and products wells. We have assumed that the driving forces for the ET process
interconverse at a rate $\gamma$ reflecting dynamic changes in the transition state.
The best tuning of the system and its highest ET rate can be achieved within the resonant
frequency region. On the other hand, nonexponential ET kinetics can be attributed to
long, time-persisting correlations in barrier configuration that effectively 
change a Poissonian character of escape events demonstrating, in general, multiscale
time-decay of initial population.   

%
%
\begin{acknowledgments}
This project has been partially supported by the Marian Smoluchowski Institute of
Physics, Jagellonian University research grant (E.G-N).

The contribution is Authors' dedication to 50th birthday anniversary of Prof. Jerzy
\L{}uczka.

\end{acknowledgments}

%
%

\end{document}